\documentclass[graphicx,preprint,showpacs,superscriptaddress]{revtex4-1}
\usepackage{amsmath,amssymb,amsthm,color}
\usepackage{graphicx}
\usepackage[english]{babel}
\usepackage{bm}

\DeclareMathOperator{\sgn}{sgn}

\begin{document}

%\preprint{}

\title{Topological characterization of Landau levels for $2$D massless Dirac fermions in $3$D layered systems}

\author{Ching-Hong Ho}
\email[]{hohohosho@gmail.com}
\affiliation{Center of General Studies, National Kaohsiung University of Science and Technology, 811 Kaohsiung, Taiwan}

\author{Ming-Fa Lin}
\email[]{mflin@mail.ncku.edu.tw}
\affiliation{Hierarchical Green-Energy Materials/quantum topology centers, National Cheng Kung University, 701 Tainan, Taiwan}

\begin{abstract}
A topological characterization of the LLs in $3$D layered systems of topological matter  is adressed. The focus is put, in a topological viewpoint, on the existence and stability of $2$D massless Dirac fermions when an external perpendicular magnetic field is applied. As is well known, it is $2$D massless Dirac fermions that provide the symmetry protection of a chiral zero-mode LL, which, in turn, dictates the half-integer QHE. The half-integer QHE is dictated by the topologically robust zero-mode LL in the relativistic LL spectrum. Two systems are considered after an introduction of some relevant topological aspects. First, $3$D layered spinless nodal-line topological semimetals (TSMs) are considered in an elaborate extent, exemplified by spinless rhombohedral graphite (RG), a $3$D stack consisting of spinless graphene layers in ABC configuration. In this system, $2$D massless Dirac fermions are hosted along the nodal lines in the bulk. Second, a discussion is held on spinful, time reversal invariant $3$D strong topological insulators (TIs), which can be layered systems such as Bi$_{2}$Se$_{3}$ and Bi$_{2}$Te$_{3}$. These systems are insulating in the bulk, while possessing an odd number of Dirac cones on the surface, where $2$D massless Dirac fermions are hosted. how to derive such an effective Hamiltonian from the lattice Hamiltonian is discussed.

\end{abstract}

\maketitle

\vskip 1.0 truecm
\par\noindent
\newpage
\section{Some topological aspects}
Topology originates as a branch of mathematics, which is concerned about classifying the spatial properties of objects with the number of holes, termed \textit{genus}, in the object surfaces\cite{seifert80}. If an object has nonzero genus, any closed path around a hole in the surface cannot be smoothly shrunk to a point. In case of that, the surface is \textit{topologically nontrivial}; otherwise, one has a \textit{topologically trivial} surface with zero genus. The central idea is the \textit{topological equivalence} between objects that have the same genus and can be mapped to each other by continuous deformation with the genus kept invariant. There is the homotopy group, which contains all the topologically equivalent objects of a certain class. The concept of topology can be naturally generalized to abstract things, even to those in the physical world. For condensed matter physics, the discovery of the integer quantum Hall effect (QHE)\cite{klitzing80} and the following topological interpretation\cite{thouless82,avron83,simon83} ought to be cited as the pioneering works. Nowadays, a new branch of physics has been built with extensive research on the topological matter\cite{kosterlitz17,haldane17,wen17}.

In the paradigmatic grand unification theory, a given thermodynamic equilibrium state of a quantum field or matter in Universe is characterized by the symmetry group (denoted $\mathcal{H}$) of, say, the Hamiltonian\cite{weinberg95}, which leads to the relevant equation for the considered system. Fundamentally, $\mathcal{H}$ is a subgroup of the largest symmetry group $\mathcal{G}$ of physical laws. Through cooling downward the ground state, $\mathcal{H}$ gets more and more reduced by successive spontaneous symmetry breaking processes. This fact puts the theoretical basis for classifying the states of physical systems in connection to phase transitions as described in, for example, the Landau's theory\cite{landau58}. That is, as symmetry is reduced, particles in a system tend to organize themselves so that the state can transit into more ordered phases, where the phase transitions occur due to spontaneous symmetry breaking in company with changes of certain local order parameter. In this approach, it is possible that there can appear topological defects\cite{mermin79}, such as solitons or domain walls, in real space of an ordered medium, like holes in the surface of an object. The determination of such phases is generally given by the nontrivial elements belonging to the homotopy group $\pi_{n}(\mathcal{H/G})$, which is defined on $S^{n}$, the $n$-sphere. Also, phase transitions of topological defects can occur with changing $\pi_{n}(\mathcal{H/G})$, as a result of spontaneous symmetry breaking.

So far, one might conclude that topology is a consequence of symmetry in the cold Universe he lives in. This is, however, not in the trend of current topological research on condensed matter. There is, indeed, an alternative approach that starts with the ground states at zero temperature, where topology is presumed to show up \textit{a priori} while symmetry is taken to be emergent\cite{volovik13}. In contrast to the reciprocal relation between local order and symmetry in the approach of the grand unification theory, the topological phases appearing here have distinguished kinds of order, which cannot be characterized by symmetry alone. Such \textit{topological order} is characterized and classified by suitable quantum numbers, namely, topological invariants\cite{wen17,wen90}. Like the genus in an object's surface, a nonzero topological invariant marks a nontrivial topological phase while a zero topological invariant marks trivial. The relevant transition between different topological phases, named topological \textit{quantum phase transition} (QPT), does not involve spontaneous symmetry breaking, whereas it occurs when the topological invariant changes at the critical point\cite{wen17,volovik07}. In the integer QHE, the topological QPT between Hall plateaus takes place with changing the first Chern number (TKNN), which is the topological invariant as given by the Hall conductivity\cite{thouless82}.

With respect to long- or short-range entanglement of wave functions in the systems considered, there could be nontrivial topological phases being robust against arbitrary or symmetry-constrained local perturbations, respectively. The phases are then divided into two respective categories: the topologically ordered phases\cite{wen17,kitaev06,levin06} and the symmetry protected topological (SPT) phases, respectively\cite{wen17,schnyder08,kitaev09,ryu10,horava05,zhao13,matsuura13,zhao16,chiu16}. For example, the fractional QHE\cite{tsui82} and high $T_{c}$ superconductors\cite{bednorz86} are nontrivial topological phases in the first category\cite{wen17}. On the other hand, there are abundant nontrivial topological phases theoretically or experimentally found in noninteracting fermionic lattice, which are largely categorized as SPT phases\cite{wen17}. An SPT phase is nontrivial as the disorder or parameter tuning respects the relevant symmetries and is only perturbative. In the single-particle band structure, the topological invariant of an SPT phase is determined with respect to the Berry phase describing the global phase evolution of the wave functions\cite{berry84}. If the bulk band is fully gapped for an insulating phase, the Berry flux over the entire Brillouin zone (BZ) is well defined\cite{xiao10}. Therefore, the topological invariant can change only when the energy gap is closed. One can think of two spatially adjacent insulators with different topological invariants in their bulk. Then, there must be a gapless state localized at the boundary between them. Since the vacuum can be taken as a trivial insulator, an SPT insulating phase of a topological insulator (TI) must be uniquely characterized by a gapless state on its surface or edge, which faces the vacuum. That is the so-called \textit{bulk-surface correspondence}, a relationship notionally invoked by the topological QPT. For an SPT phase, the surface states are protected by the bulk symmetry and take a form depending on the symmetry\cite{hatsugai93,ryu02,qi06,hatsugai09,mong11}.

Also, the notion of topological QPT may lead to the emergence of topological semimetals (TSMs) in Nature. Recall the thought underlined in the bulk-surface correspondence; now again, think of two spatially adjacent systems, with one of them being a TI while the other, a TSM. This two-system picture can be identified as a three-system picture, where the TSM is intervened between the TI and the vacuum (a trivial insulator). Thus, the possibility of a TSM mediating between two insulators is expected. In general, a TSM has band crossing between bands with the same symmetry, as opposed to the avoided level crossing stated in the von Neumann-Wigner's theorem for trivial semimetals\cite{demkov07}. Here a node can be identified where the band crossing (almost) is generated at the Fermi level. To described the nodes, a condition, say, for a two-band model, is given by $d_{f}-d_{n}+c_{s}=2\times2-1=3$ with $d_{f}$ the degrees of freedom, $d_{n}$ the nodes' dimension and $c_{s}$ the the number of symmetry constraints. It is noted that $d_{f}$ depends on how the nodes are constrained. If the maximum $c_{s}=3$ is required, the one acquire $d_{n}=d_{f}$ so that the nodes are fixed at the symmetry points or lines. These so-called \textit{essential} band crossings are completely protected by symmetries, which are crucially nonsymmorphic group symmetries, possibly together with nonspatial symmetries\cite{young12,young15,yang17}, and are not considered here. On the other hand, in case that $c_{s}<3$, the band crossing is \textit{accidental}. The significance of \textit{accidental} nodes was issued in the early days for time reversal invariant crystals\cite{herring37}. Nodes at \textit{accidental} band crossings acquire degrees of freedom to adapt to perturbative parameter tuning, which preserves the required symmetries, to be moved over generic points in the BZ. Their stability depends on if they are resided in the interior of a finite and connected region in the parameter space, and they are unstable, or vanishingly improbable, at the margin of the region. That is, the nodes are only perturbatively stable and they can be removed when the deformations are too large. Return to the topological QPT. Now, for the phase transition from an SPT insulating phase to a trivial insulating phase, at the critical point the band gap should be closed somewhere in the BZ via \textit{accidental} band crossing, which is marginal\cite{murakami07,murakami08}. Inspired by the two-system picture thought above, one would expect the possibility of a TI-to-TSM transition, which amounts to broadening the margin between TI-to-trivial insulator so as to have a region in the parameter space for \textit{accidental} band crossing to reside inside. These stable nodes give rise to a TSM\cite{yang13,murakami17}.

In comparison to the topologically ordered phases, which are robust against arbitrary perturbation, the SPT phases are, in a sense, less robust since they need the protection of certain symmetry. The perturbation an SPT phase can survive is limited to preserve the symmetry specific to the very phase. Hence, it is of fundamental importance to characterize and classify the SPT phases of condensed matter. For this purpose, one should know that the severest perturbation comes from disorder. Since early days, physicist have known that in a noninteracting electronic system in spatial dimension $d\leq2$, the ground-state wave function (at zero temperature) can be localized by disorder in the thermodynamic limit\cite{anderson58}. This so-called \textit{Anderson localization} stands as the basis for the QPT. The advent of the scaling theory introduced the universal classification of the localization behavior for noninteracting fermionic systems in accordance to symmetry and spatial dimensionality\cite{abrahams79,anderson80}. This classification was solidly expressed later by using the random matrix theory\cite{zirnbauer96,altland97}. After including the new class of anti-localization found in the time reversal invariant TIs and topological superconductors (TSCs), a \textit{tenfold way} classification was established for noninteracting insulating systems, where ten universal symmetry classes are given, with characteristics depending on the spatial dimensionality in a period-$8$ periodic table\cite{schnyder08,evers08}. The ten symmetry classes comprise the combination of three nonspatial symmetries with regard to the presence or absence of electron spin $SU(2)$ rotation symmetry. Spatial symmetries are excluded from the \textit{tenfold way} classification because they are prone to disorder. The three nonspatial symmetries are time-reversal symmetry (TRS), particle-hole symmetry (PHS) and sublattice symmetry (SLS). It is noted that in the context of the random matrix theory, SLS is alternatively called chiral symmetry (CS) owing to that the two sublattices lead to a spinor. Both TRS and PHS are antiunitary while SLS is unitary. Of them, any one can be derived from the product of the other two. Such a way of classification can be proved mathematically using the $K$ theory\cite{kitaev09}.

The situation of SPT semimetallic phases is more involved because of their nodal band structures, where the nodes are pinned to or close to the Fermi energy $E_{F}=0$. A node leads to singularity in the BZ and, therefore, one cannot derive a topological invariant by integration over the whole BZ. Thus, another \textit{tenfold way} classification has also been devoted to semimetallic phases with nodes of dimension $d_{n}$ at \textit{accidental} band crossings\cite{horava05,zhao13,matsuura13,chiu14,zhao16}, where the codimension $p=d-d_{n}$ is taken into account. This classification is defined for a submanifold that is gapped around the node so that the topological invariant is acquired in the submanifold. Such nodes are stable individually, to the extent of symmetry-preserving disorder or perturbation of parameter tuning, when they are characterized by nonzero topological invariants. In contrast to the case of SPT insulating phases, the protection of nodes in the BZ often is achieved by the nonspatial symmetries together or in combination with certain spatial symmetries\cite{chiu14,koshino14}. In the continuum limit in the vicinity of a single node between two bands the quasiparticles show up as two-component fermions, which are, however, forbidden in a lattice by the Nielsen-Ninomiya's no-go theorem\cite{nielsen81a,nielsen81b,nielsen81c}. The theorem in the framework of topological condensed matter states that in a local-action, real and noninteracting fermionic lattice, the number of species of existing fermions must be double with opposite topological charges, say, chiralities. That is, there must be as many fermions of positive chirality as of negative chirality and consequently the components of the fermionic wave function are doubled. One thus obtain four-component Dirac fermions, instead of two-component fermions. This \textit{fermion doubling} problem arises when a continuous field are discretized into a lattice. It has a topological origin. Due to the opposite topological charges of the nodes existing in pair, the Berry phase obtained by integration along the border of the BZ must vanish. Hence, the fate of node annihilation cannot be got rid of as the nodes at accidental band crossing are moved toward each other under symmetry-preserving disorder or perturbation. Around linearly crossing nodes, i.e., Dirac points (DPs), the Dirac fermions hosted are massless. Massless Dirac particles are ubiquitous in condensed matter, for they obey certain loosely identified Dirac equations\cite{vafek14} as compared to the genuine Dirac equation in the relativistic quantum mechanics\cite{dirac58}. In this context the Lorentz invariance has dropped out from the mimic Dirac equations. Moreover, the components of wave functions need not comply with the requirement of the genuine Dirac equation ($2^{d}$ for $d=2,3$); rather, it depends on the number of spinors adapted to the system under consideration. Those spins are acquired \textit{a priori} in contrast to the electron spin, the latter being intrinsically derived from the genuine Dirac equation. In the forbidden case of two-component Weyl fermions around a single nodes, the Hamiltonian is naively casted as $H(\mathbf{p})=v\mathbf{\sigma}\cdot\mathbf{p}$, with $v$ the Fermi velocity and $\sigma$ the Pauli matrices acting on the spinor. Obviously, the eigenstates of $H(\mathbf{p})$ are chiral since they are simultaneous eigenstates of the helicity operator $h=\mathbf{\sigma}\cdot\mathbf{p}/|\mathbf{p}|$\cite{kitaev09,ryu10}. This means that $H(\mathbf{p})$ has a CS effectively derived in $d$ dimensional space, differing from the CS defined in $(d+1)$ dimensional Minkowski spacetime.

The first demonstration of $2$D massless Dirac fermions in condensed matter was addressed on the $2$D honeycomb lattice\cite{semenoff84}, where spinless condition or spin degeneracy ($SU(2)$ symmetry) was assumed while the two sublattices come into play as a pseudospin. With only the nearest neighbor tight binding hoppings (minimal model), there would exist two DPs in the valleys around which $2$D massless Dirac fermions of chirality $+1$ and chirality $-1$ are separately hosted. This leads to another pseudospinor\cite{hatsugai11}. Hence, four-component $2$D massless Dirac fermions have been recognized to be promising in an analog to the $(2+1)$-dimensional gauge field. The possibility has been proposed in the $2$D limit of graphite, which was shown to be stable against impurity without intervalley scattering\cite{divincenzo84}. As was later realized and extensively studied\cite{castroneto09}, these results has been well realized from spinless graphene. On the basis of these two pseudospinors, the Dirac Hamiltonian can be derived in the continuum limit around the two DPs. Within the minimal model, it is given by $H(\mathbf{k})=v\hbar\mathbf{\tau}_{3}\mathbf{\sigma}\mathbf{k}$, with $\hbar \mathbf{k}$ the crystal momentum measured from each DPs, $v$ the Fermi velocity defined by the nearest neighbor lattice site distance and hopping integral, and $\mathbf{\sigma}$ and $\mathbf{\tau}$ the Pauli matrices acting on the sublattice and the valley spinor, respectively. In this Dirac Hamiltonian, intervalley coupling is absent. This Dirac Hamiltonian carries CS if intervalley is absent, as a result of SLS present in the relevant lattice Hamiltonian\cite{hatsugai11}. The CS of the Dirac Hamiltonian is termed continuous CS in the following. With $\lambda$ and $\xi$ denoting the band and valley indices, respectively, the energy of Dirac fermions is given by $E_{\lambda \xi}(\mathbf{k})=\lambda v\hbar \mathbf{k}$. The wave function is given by
$
\psi_{\lambda\xi}(k)=(1/\sqrt{2})\left(\exp{(-i\xi\theta/2)},
\lambda\xi\exp{(i\xi\theta/2)}
\right)^{T},
$
where $\mathbf{k}=(k, \theta)$ has a phase defined conventionally. It is easy to show that a winding number characterizing the CS class of the Dirac Hamiltonian is given by $w_{1}=\lambda \xi$ with respect to the chirality. In the presence of a uniform, perpendicular magnetic field $B\hat{\mathbf{z}}$, continuous CS is still preserved via the minimal coupling. This fact is simply manifested in clean graphene. The Landau level (LL) energy spectrum is obtained as $\epsilon_{n}=\sgn{(n)}\hbar \omega_{c} \sqrt{|n|}$, with $\omega_{c}=v\sqrt{2eB/\hbar c}$, where the $\sqrt{B}$ function reflects the linear dispersion. The LL energy spectrum is characterized by a zero mode that is constantly pinned to zero energy in a wide range of magnetic field as revealed in the Hofstadter butterfly\cite{hofstadter76}. The LL wave functions are given by
$
\Psi(\mathbf{r})\propto\left(
\phi_{|n|},
\sgn(n)\phi_{|n|-1}
\right)^{T},
$
with $\phi_{|n|}$ being the $n$th simple harmonic oscillator wave function, so that the zero-mode LL is pseudospin polarized. Therefore, the zero-mode LL is still chiral and half-filled with respect to the DPs. It is robust under the protection of the continuous CS\cite{hatsugai11,kawarabayashi10a,kawarabayashi10b}. It is remarkable that the chiral zero-mode LL can serve as a mark of $2$D massless Dirac fermions existing in a noninteracting fermionic lattice with SLS and the resulting continuous CS. With the zero-mode LL, the half-integer quantized Hall conductivity $\sigma_{xy}=4(n+1/2)e^{2}/h$ has been inferred in the context of spinless graphene in the quantum Hall limit\cite{zheng02,gusynin05}, which was realized soon later\cite{novoselov05,zhangy05}. It has been understood that the half-integer QHE is dictated by the chiral, half-filled zero-mode LL\cite{kawarabayashi09,kawarabayashi10a}, which is ascribed to continuous CS of $2$D massless Dirac fermions. Hence, disorder that degrades continuous CS would wash away such a characteristic and makes the half-integer QHE impossible. It should be noted that the once upon a transition between Hall plateaus, a topological QPT takes place as the chemical potential passes through a LL where the criticality should be characterized\cite{evers08,laughlin81,halperin82}. A significant characteristic of continuous CS is that it anomalously dominates the criticality at the zero-mode LL\cite{altland97,kawarabayashi09,ludwig94,hatsugai97}, which is preserved by random happoins, irrespective of that the Hall plateaus are assured by existing localization states due to disorder in ordinary integer QHE\cite{ostrovsky08}.

The present concern is about the topological characteristics of $2$D massless Dirac fermions that are hosted in the vicinity of nodes in $3$D layered systems. Generally speaking, it is easy to figure out that there are only three types of nodes with massless Dirac fermions in $3$D noninteracting lattice systems\cite{volovik03}. In the first type shown in Fig. 11.1(a), $3$D massless Dirac fermions are hosted around isolated DPs in the bulk, where the Dirac cones exhibit linear dispersion in all the three dimensions. The notable Weyl semimetals\cite{hosur10,wan11,burkov11b,halasz12,fang12} and Dirac semimetals \cite{wang12,liu14} are of this type and are excluded. Intuitively, the possibility for $2$D massless Dirac fermions existing in $3$D is rendered by the rest two types. In the second type, $2$D massless Dirac fermions exist in codimension $p=2$ around Dirac nodal lines (DNLs) in the bulk of a $3$D BZ, as shown in Fig. 11.1(b). There are infinitely many Dirac cones along a DNL, each dispersing linearly in the the $2$D plane of its own. The stable existence of DNLs indicates a nodal-line TSM. On each of the surface layers, the surface state manifests itself by a drumhead flat band corresponding to each bulk DNL. The simplest model is, among others, the rhombohedral lattice, which is the rhombohedral stack of honeycomb-lattice layers. The physical realization has been found from spinless rhombohedral graphite (RG)\cite{xiao11,heikkila11,ho13,ho14,heikkila15,ho16}. The first and second types comprise TSMs inclusively\cite{burkov11a,yang14,armitage18}. On the other hand, the third type includes those having fully gapped bands in the bulk and corresponding surface Dirac cones, where $2$D massless Dirac fermions are hosted on each surface, as shown in Fig. 11.1(c). This type indicates $3$D time reversal invariant TIs\cite{murakami07,murakami08,fu07a,fu07b}.

At final, the special character of $3$D layered systems is remarked. When identical layers of $2$D noninteracting fermionic lattice are stacked into $3$D, the tight binding interlayer hoppings, whether strong or weak, would bring the stack into a topological class that is either the same as or different from a single layer. People usually tend to use a model in the $2$D limit for a $3$D layered system provided that the interlayer coupling is weak compared to certain attributes of the $2$D lattice. However, the quasi-$2$D model does not work always. The most famous example is the spinless $2$D graphite sheet (nowadays, graphene), which were for modelling graphite\cite{wallace47}. It has been known, from the magneto-electronic and magneto-optic properties\cite{ho13,ho14,mcclure69,guinea06,ho15}, that this quasi-$2$D model fails for Bernal graphite but reveals mimic results for RG. Other examples have been found in some organic conducting salt, like $\alpha$-(BEDT-TTF)$_{2}$I$_{3}$, or strained graphene\cite{goerbig08}, in which interlayer magneto-resistance plays a role though the quasi-$2$D model is frequently employed. The success of the quasi-$2$D model for spinless RG suggests the topological characteristic inherited. When a $2$D layer of lattice possesses node points in the bulk, whether it being realistic material or being an idealized model, one might ask of what if the layers are stacked to $3$D in some ways. For systems having a quasi-$2$D used for granted, such as d-wave cuprate superconductor with negligible interlayer coupling\cite{anderson87}, one would naively derive vertical nodal lines from the nodal points of each layer\cite{chiu14}. However, in stacks of graphene layers, where the interlayer hoppings are significant, the results could disperse in the stacking dimension in various manner depending on the stacking configurations\cite{dresselhaus02}. Bernal (AB stacked) graphite has a nonsymmorphic lattice\cite{dresselhaus08}. In such a system, nexuses can show up where nodal lines merge\cite{heikkila15,hyart16}. By contrast, in spite of that DPs are known to exist in the $\pi$-flux square lattice, chiral TIs and Weyl semimetals can be obtained by stacking layers of the lattice into a specific $\pi$-flux cubic lattice\cite{hosur10}. All these are examples of the dimensional crossover from $2$D to $3$D\cite{volovik07}. The (co)dimensional-periodic table in the \textit{tenfold way} classification dictates that the topological class would or would not alter after the dimensional crossover. RG is a $3$D layered system consisting of graphene layers stacked in rhombohedral (ABC) configuration. From graphene to RG, the topological phase changes from a $2$D TSM with DPs to a $3$D TSM with DNLs in a dimensional crossover\cite{heikkila11,ho16}, where the codimension $p=2$ holds. As having illustrated [Fig. 11.1(b)], the DNLs in spinless RG fall into the second type of nodes, around which $2$D massless Dirac fermions are hosted as well as in spinless graphene. In the presence of a perpendicular magnetic field, a $3$D layered system has the easiest way to satisfy the Diophantine equation for LL gaps opened in the $3$D bulk, as is required for the $3$D integer QHE\cite{kohmoto92}. It is interesting to compare the $2$D integer QHE of a single layer to the $3$D integer QHE of the layered system\cite{balents96}. A comparison between graphene and RG is of particular interest, where $2$D massless Dirac fermions hosted in both systems are responsible for the existence of a robust zero-mode LL with respect to the the Dirac nodes, which, in turn, is a necessary condition of the half-integer QHE. The $3$D half-integer QHE held has been experimentally confirmed\cite{kopelevich03,kempa06}. A schematic plot of $3$D integer QHE in a $3$D layered system is provided in Fig. 11.2.

\section{Chiral symmetry protected nodal-line topological semimetals}
Here, RG as a $3$D layered system consisting of graphene layers is introduced, with the assumption of spinless condition or $SU(2)$ symmetry. This system has been well analyzed but less topologically characterized in spite of the well understanding of graphene. Spinless graphene is not only deemed a realistic system with negligible spin orbital coupling (SOC) or a toy model, but also anticipated to be an artifact on the honeycomb lattice\cite{wunsch08} as being realizable by means of cold atoms in optical lattices\cite{chen12}, where the tight-binding hoppings are included to adapt to the model. The honeycomb lattice is symmorphic\cite{dresselhaus08}, for which the relevant point-group symmetries are set forth as follows\cite{asano11}, referring to Fig. 11.3(b). There is a $C_{3}$ rotation (by $2\pi /3$) symmetry at each of the A (B) sublattice sites. Also, there is a mirror reflection symmetry (MRS) with the reflection axis put between A and B; yet, another one, between A (B) and A (B), is present as well but does not play an important role for protecting the DPs. It is the $C_{3}$ rotation symmetry together with certain symmetries, if being preserved in the model Hamiltonian, that makes the two DPs fixed at the $K$ and $K^{\prime}$ points of the BZ\cite{koshino14,asano11}. This \textit{essential} band crossing is only limited, not required for the existence and stability of DPs at more generic momenta with \textit{accidental} band crossing\cite{hasegawa06}. For the model including up to the next-nearest neighbor hoppings, TRS together with MRS can be present and protects the DPs\cite{chiu14,wunsch08}, which are located on the reflection axis. This SPT semimetallic phase belongs to the symmetry class AI with MRS. The associated topological invariant is given by the integer $Z_{R^{\pm}}$ ascribed to the eigenvalues of MRS, where TRS is respected in both the eigenspaces\cite{chiu14}. This model could also render DPs each pretected by spacetime inversion symmetry (PTS), i.e., the combination of TRS and space inversion symmetry (IS), which commutes with the Hamiltonian, i.e., $[\mathcal{H},\mathcal{T}\mathcal{I}]=0$\cite{manes07,raghu08}. The relevant symmetry class also is AI, but now the topological invariant is given by the binary number $Z_{2}$ for each DP because $\mathcal{T}\mathcal{I}$ transforms a DP into itself in the BZ. This fact can be described in spatial dimension $d=7$ in the periodic table in the \textit{tenfold way} classification, as if the dimension of the submanifold ($S^{1}$ here) around each DP were $-1$\cite{teo10,morimoto14}. It is noted that PTS can protect DPs at generic momenta off the the reflection axis in the case of generally anisotropic hoppings due to randon hopping, say, while TRS together with MRS provides the protection up to the case of the hoppings do not break MRS\cite{hasegawa06}.

If the model for spinless graphene is reduced to merely include the nearest neighbor hoppings, there is SLS additionally, which coexists with PTS. In this model, SLS is preserved under the presence of IS as a necessary condition. Hence, SLS should be respected in each eigenspace of PTS\cite{koshino14}. As mentioned above, SLS behaves as CS so that continuous CS in $S^{1}$ around each DP is present in the absence of intervalley coupling. The SLS operator anticommutes with the Hamiltonian, i.e., $\mathcal{S}\mathcal{H}\mathcal{S}^{-1}=-\mathcal{H}$, and transforms an eigenstate to its conjugate that has inversed eigenenergy and an eigenfunction having one of the two components inversed on the two sublattices, i.e., $\mathcal{S}\psi_{E}=\psi_{-E}$ with $\psi_{E}=(a,b)^{T}$ and $\psi_{-E}=(a,-b)^{T}$\cite{hatsugai11,zhao13}. This minimal model also belongs to an additional symmetry class BDI, where the present SLS comes out to be the combination of TRS and PHS. The energy symmetry between particles and holes is held by SLS as well as PHS. In the momentum representation, the topological invariant derived from the Berry phase is now identical to the winding number $w\in Z$\cite{zhao13,koshino14}, which is defined in each $S^{1}$ submanifold\cite{wen89}. The bulk-edge correspondence can be deduced from a knowledge of the winding number, manifesting itself by flat bands pinned to the Fermi energy $E_{F}=0$ on the zigzag edges\cite{ryu02,hatsugai09}. Moreover, as a result of SLS, the chiral zero-mode LL is protected by continuous CS and pinned to $E_{F}=0$ as well\cite{hatsugai11,kawarabayashi10b}. Beyond the minimal model, the DPs are protected by PTS only. Inclusion of the next-nearest neighbor hoppings destroys SLS and introduces significant effects. In this model, the DPs shift away from the Fermi energy and the surface bands change from being flat to being linear; besides, the Dirac cones become tilted where tilted $2$D massless Dirac fermions are hosted, similar to certain organic conducting salt\cite{goerbig08}. Nevertheless, it is remarkable that tilted massless $2$D Dirac fermions still has a generalized continuous CS so that the zero-mode LL is protected\cite{kawarabayashi11,kawarabayashi16}.

The lattice of spinless RG, as shown in Fig.s 11.3(a) and 11.3(c), is the $3$D  rhombohedral (ABC) stack (along the $z$ direction) of $2$D honeycomb-lattice layers [coordinated by $(x,y)$]. Obviously, the lattice consists of two sublattices as well as the honeycomb lattice does; also, it is symmorphic\cite{dresselhaus08}. Regarding the point group symmetries, it is easy to know that there is no any rotation symmetry about the hexagonal $K$ or $K^{\prime}$ lines, referring to Fig. 11. 4(b). Moreover, MRS is absent from this lattice while IS is present. The lattice Hamiltonian including up to the next-nearest neighbor hoppings follows as $\mathcal{H}=\beta_{0}\sum_{l}\sum_{\langle i,j\rangle}[a_{l,i}^{\dag}b_{l,j}+\mathrm{h.c.}]%
+\beta_{1}\sum_{l}\sum_{i}[b_{l+1,i}^{\dag}a_{l,i}+\mathrm{h.c.}]%
+\beta_{4}\sum_{l}\sum_{\langle i,j\rangle}[a_{l+1,i}^{\dag}a_{l,j}+b_{l+1,i}^{\dag}b_{l,j}+\mathrm{h.c.}],
$
with $a_{l,i}^{\dag}$ ($b_{l,i}^{\dag}$) create fermions in the sublattice $A$ ($B$) at site $i$ on layer $l$, where $l$ labels the layers and $\langle i,j\rangle$ denotes both the nearest neighbor intralayer and the next-nearest neighbor interlayer hoppings. Besides the nearest neighbor intralayer ($\beta_{0}$) and interlayer ($\beta_{1}$) hoppings, the next-nearest neighbor interlayer ($\beta_{4}$) hoppings are included optionally. Under the Fourier transformation $a_{l,i}=N^{-\frac{1}{2}}\sum_{\mathbf{k}}e^{-i\mathbf{k}\cdot\mathbf{r}_{i,l}}a_{\mathbf{k}}$,
(similar for $b_{\mathbf{k}}$), the Bloch Hamiltonian is written as
$H(\mathbf{k})=\sum_{i=0,1,2}d_{i}(\mathbf{k})\sigma_{i}$,
with the Pauli matrices acting on the space spanned by the two sublattices $(a_\mathbf{k},b_\mathbf{k})$, where $d_{1}(\mathbf{k})=-\beta_{0}\sum_{m}^{3}\cos{(\mathbf{k}_{||}\cdot\mathbf{\delta}_{m})}%
+\beta_{1}\cos{(k_{z}c)}$, $d_{2}(\mathbf{k})=\beta_{0}\sum_{m}^{3}\sin{(\mathbf{k}_{||}\cdot\mathbf{\delta}_{m})}%
-\beta_{1}\sin{(k_{z}c)}$ and $d_{0}(\mathbf{k})=2\beta_{4}\sum_{m}^{3}\cos{(\mathbf{k}_{||}\cdot\mathbf{\delta}_{m}-k_{z}c)}$ are obtained with $\mathbf{\delta}_{m}$ the three vectors connecting nearest neighbor lattice sites in a layer and $c$ the layer distance.

The symmetries of the Bloch Hamiltonian $H(\mathbf{k})$ is described as follows. At first, IS is preserved as $\mathcal{I}H(\mathbf{k})\mathcal{I}^{-1}=H(\mathbf{-k})$, with the operator $\mathcal{I}=\sigma_{1}$\cite{manes07}. Regarding the nonspatial symmetries, TRS is preserved as $\mathcal{T}H(\mathbf{k})\mathcal{T}^{-1}=H(\mathbf{-k})$, with the operator $\mathcal{T}=\sigma_{0}\mathcal{K}$, where $\sigma_{0}$ is the $2\times2$ identity matrix and $\mathcal{K}$ is the complex conjugation operator. In the presence of TRS, SLS and PHS are preserved or not in company. For SLS with $\mathcal{S}=\sigma_{3}$, it is preserved as $\mathcal{S}H(\mathbf{k})\mathcal{S}^{-1}=-H(\mathbf{k})$ if $\beta_{4}=0$ (minimal model) while being broken if $\beta_{4}\neq0$. It is easy to verify the situation of PHS with the operator given by $\mathcal{C}=\sigma_{3}\mathcal{K}$. Therefore, the model for spinless RG including the next-nearest neighbor hoppings belongs to the symmetry class AI, in terms of the \textit{tenfold way} classification. This class, however, does not allow any nontrivial nodes at accidental band crossing, whether the codimension $p=2$ or $3$\cite{chiu14}. The protection of nodes in this $3$D layered system needs additionally certain spatial symmetries. Indeed, there is a large category of spinless nodal-line TSMs that is protected by PTS, i.e., $[\mathcal{H},\mathcal{T}\mathcal{I}]=0$\cite{kim15,fang15,chan16}, and spinless RG falls into this category. It can be shown that, given PTS , DNLs can stably exist in spinless RG under the protection of nonzero $Z_{2}$ topological invariant\cite{volovik13}. When SLS and PHS is recovered to coexists with PTS within the minimal model, the system belongs to an additional class BDI. Now, the winding number $w_{1}=(2\pi i)^{-1}\triangledown_{\mathbf{q}}\log \sigma_{3} H(\mathbf{k})$ can be defined in each $S^{1}$ submanifold in momentum representation\cite{heikkila11}.

To determine the existing nodes in spinless RG, here one has three variables ($k_{x}$, $k_{y}$, $k_{z}$) and two equations ($d_{1}=d_{2}=0$) from the Hamiltonian $H(\mathbf{k})$. Hence, the solution generally manifests itself by lines in $3$D $k$ space. The very character of the present layered system is disclosed by not only the typical values of the hopping integrals, $\beta_{0}\gg\beta_{1}\gg\beta_{4}$, but also the form of $H(\mathbf{k})$. It guides one to find the zeros around the hexagonal $K$ and $K^{\prime}$ lines with respect to graphene\cite{mcclure69,ho13}. As expected, a pair of DNLs has been found as shown in Fig. 11. 4(b), which are almost exactly expressed by $\hbar k_{DL}=\beta_{0}/v_{0}$ and $\phi_{DL}=\xi k_{z}c$ in polar coordinates $\mathbf{k}=(k,\phi)$ with respect to the $K$ and $K^{\prime}$ lines, where $v_{0}=3a\beta_{0}(2\hbar)^{-1}$ is defined for a graphene layer with $a$ the nearest neighbor site distance and $\xi =\pm1$ respectively denote the two DNLs. The two DNLs appear to spiral around the $K$ and $K^{\prime}$ lines respectively in opposite senses across the BZ boundaries from $k_{z}=-\pi$ to $k_{z}=\pi$. One can carry out a coordinate transformation in terms of $(q, \theta)$ measured from the DNLs at constant $k_{z}$ plane\cite{ho13}. Consequently, two tilted Dirac cones stand at the DNLs for constant $k_{z}$, given by $\varepsilon(q, \theta)=\lambda[2 v_{4}\cos{(\theta+\xi k_{z}c)}+ \xi v_{0}]\hbar q$, with $v_{4}=3a\beta_{4}(2\hbar)^{-1}$, where $\lambda=\pm$ are the bands indices. The wave functions of spinless RG have also been known\cite{ho14}, which were shown to be almost chiral as those of spinless graphene, to next higher order of $\beta_{4}$ while being independent of $\beta_{1}$. Thus, there are tilted $2$D massless Dirac fermions hosted around the DNLs in spinless RG.

The Dirac cones in spinless RG become normal in the minimal model with $\beta_{4}$ vanishing, so that the stacking dimension $k_{z}$ completely drops out. It has been proven that, with this, RG mimics graphene in every aspect, including the density of states \cite{guinea06,ho16}, magneto-electronic properties such as LLs with a zero mode\cite{guinea06,ho13,ho14} and magneto-optic properties\cite{ho15}. It is reasonable since spinless RG bearing SLS hosts $2$D massless Dirac fermions around the two DNLs. In the more realistic model, the inclusion of $\beta_{4}$ brings out certain modifications in regard to the SLS breaking. That is, the DNLs does not lie at $E_{F}=0$ and the topologically corresponding drumhead surface bands would deviate from being exactly flat. Those modifications have been shown in previous experiments to be negligible, not yet topologically characterized, such that almost exactly flat surface bands were shown in epitaxied RG\cite{pierucci15} and the $3$D half-integer QHE was shown in natural graphite\cite{kopelevich03,kempa06}. In spite that the DNLs for tilted $2$D massless Dirac fermions in spinless RG are protected by PTS and characterized by the $Z_{2}$ invariant, there remains an issue that is crucial in the topological characterization of LLs. Specifically, it is desirable to determine whether a stable zero-mode LL exists, which would be responsible for the $3$D half-integer QHE. As numerical results have shown\cite{ho13,ho14}, the LL spectrum (Fig. 11. 5) and LL wave functions (Fig. 11. 6) for spinless RG with $\beta_{4}\neq0$ can be identified to mimic those of spinless graphene except for the $k_{z}$ dispersion in the LL spectrum. However, the model system lacks continuous CS for protecting the zero-mode LL since SLS is absent due to nonvanishing, though small, $\beta_{4}$. Here, a generalized continuous CS operator anticommuting with the Hamiltonian $H(\mathbf{k})$ is proposed, given by $\Gamma=\rho^{-1}[\sigma_{3}-i\eta(\sin{(\xi k_{z}c)}\sigma_{1}+\cos{(\xi k_{z}c)}\sigma_{2})]$, with $\rho^{-1}=[1-(\beta_{4}/\beta_{0})^{2}]^{1/2}$. In a similar manner to other systems that have tilted $2$D massless Dirac fermions hosted, the generalized continuous CS is respected so that the zero-mode LL is protected.

\section{Time reversal symmetry protected $3$D strong topological insulators}
Graphene, again! For the distinguished topological phases of time reversal invariant TIs, it is graphene that invoked the first notification\cite{kane05a,kane05b,hasan10}. The TRS protection makes this kind of TIs differing from those Chern insulators, e.g., integer quantum Hall insulators, whose topological invariant, e.g., Hall conductivity, is odd under time reversal operation. With SOC, SLS and IS are broken. A mass term, though being tiny, is then induced so as to gap the bulk bands of graphene. In the relevant Hamiltonian $H(\mathbf{p})=vs_{3}\mathbf{\tau}_{3}\mathbf{\sigma}\mathbf{p}$, an additional spinor acting on the electron spin space is included. In comparison to integer quantum Hall insulators, which have chiral edge states carrying electrical current, TRS protected TIs have helical edge states carrying carry spin current and exhibiting the quantum spin Hall effect (QSHE). This bulk-edge correspondence can be proved solidly through the Kramer degeneracy derived from TRS\cite{qi06}. Because of TRS, such a nontrivial topological phase is characterized by nonzero $Z_{2}$\cite{moore07}. The experimental realizations were achieved soon later by means of HgTe/CdTe quantum wells that have much larger SOC gap\cite{bernevig06,konig07}.

Still, the $3$D QSHE is possible if $2$D layers of quantum spin Hall insulator are stacked to $3$D in a specific way\cite{ran09,slager14}, while keeping the QSHE on each layer in analog to the $3$D integer QHE\cite{kohmoto92}. The resulting $3$D layered system is a kind of $3$D weak TI. There exist an even number of Dirac cones on each surface layer. This might be a manifestation of the quasi-$2$D character of $3$D TRS protected weak TIs, in which each layer has fermion doubling. In the \textit{tenfold way} classification\cite{schnyder08}, these systems belong to the symmetry class AII, to which the $Z_{2}$ is specified.

On the other hand, TRS protected $3$D strong TIs are found as having been predicted\cite{fu07a,fu07b}. They also are layered systems as realized by Bi$_{2}$Se$_{3}$, Bi$_{2}$Te$_{3}$, etc, which have the stacking units consisting of one quintuple layer sandwiched by Bi$_{2}$ and Se$_{3}$ ( Te$_{3}$) in a rhombohedral configuration\cite{hsieh08,chen09}. These TIs exhibit strong SOC in an inverted band structure as required. Under the protection of TRS, $3$D strong TIs belong to the symmetry class AII and are characterized by $Z_{2}$. By contrast, there exist a single or an odd number of Dirac cones on each surface\cite{hsieh08,chen09}, corresponding to the gapped bulk bands\cite{qi06}. The Kramer degeneracy forces the DP be located at the time reversal invariant point on each surface BZ. The existing $2$D massless Dirac fermions on the surfaces is attributed to the third type of nodes in $3$D, as described in Fig. 11.1(c). Because the two surface layers are practically distant from each other and, hence, their hybridization can be neglected; that is, the intervalley mixing is circumvented. Moreover, the Dirac fermions can penetrate into the bulk hardly. Hence, it is reasonable to consider the single DP on an individual surface. TRS also leads to an exotic helical spin texture, which has been experimentally observed\cite{kung17}. In this texture, electron spin is always locked to momentum, as shown in Fig. 11.7. Thus, there is spin polarization away from the DP, rather than spin degeneracy as in graphene. To sum up, the TRS protected strong TI host one fourth massless Dirac fermions with only one spinor in comparison to spinless graphene.

To characterize the single DP on a surface of TRS protected $3$D strong TI, one might require an effective Dirac Hamiltonian $H(\mathbf{p})=v\mathbf{\sigma}\bf{p}$, with $v$ the Fermi velocity and $\sigma$ the Pauli matrices acting on the electron spin space. Of course, a Dirac Hamiltonian acquires a continuous CS. This seems to be intriguing since the system lacks SLS. The cause should be attributed to the Kramer degeneracy, again, since there is no symmetry constraint else. However, how to derive such an effective Hamiltonian from the lattice Hamiltonian is crucial. In the conventional methodology, it is proper to get a surface Hamiltonian from the finite lattice Hamiltonian since both the bulk and the surface are non-local to each other. This is why fermion doubling is absent from the surface of the TRS protected $3$D strong TI. However, in so doing some perturbation terms would be brought out, which can degrade the continuous CS and, therefore, gap the DP. This problem should be reconciled in order to cast a Dirac Hamiltonian to characterize the $2$D massless Dirac fermions on the surfaces of the system\cite{deresende17}.

As well realized, the surface Dirac cone is gapped by the exchange field due to magnetic doping or a proximate magnetic material\cite{eremeev13,yoshimi15a}. Referring to Fig. 11. 8(c), the LLs lose the characteristic of $2$D massless Dirac fermions. QHE in this case is anomalous [Fig. 11. 8(d)]. However, all experiments till now have shown that applying a magnetic field perpendicular to the surface can lead to a chiral, half-filled zero-mode LL [Fig. 11. 8(a)], which is ascribed to $2$D massless Dirac fermions\cite{cheng10,hanaguri10}. The half-integer QHE has also been observed\cite{xu14,yoshimi15b}, referring to [Fig. 11. 8(b)]. The LL degeneracy is one fourth of those in graphene. Remember, however, the continuous CS is preserved via the minimal coupling. Thus, one would obtain an anomalous QH effect.

\newpage
\bigskip \vskip0.6 truecm\newpage

\bigskip \vskip0.6 truecm\newpage
\centerline {\Large \textbf {Figure Captions}}
\begin{itemize}

\item[FIG. 11.1] Types of nodes in the $3$D BZ. (a) Nodal points inside the bulk. (b) Two types of nodal lines inside the bulk. The line can be an interior loop or a periodic line across the BZ boundaries. (c) Nodal points on the $2$D surface BZ.

\item[FIG. 11.2] Schematic plot for $3$D QH effect with spin current along the chiral edges or for $3$D weak TI with spin current along the helical edges.

\item[FIG. 11.3] (a) Rhombohedral lattice of $3$D stack of $2$D layers of honeycomb lattice, where the rhombohedron (red) is the biparticle primitive unit cell. The present minimal model is described in terms of the intralayer hopping $t$ and interlayer hopping $t^{\prime}$. The two sublattices are respectively given by solid and open dots. (b) Honeycomb lattice of each $2$D layer. The nearest-neighbor sites associated with hopping $t$ are connected by three vectors $\mathbf{\delta}_{m}$. The two sublattices are respectively given by solid and open dots. (c) Schematic of the $3$D extension of SSH model constructed by a rhombohedral stack of $2$D honeycomb-lattice layers. One representative chain is shown by linked thick sticks where intralayer hopping $t$ (blue) and interlayer hopping $t^{\prime}$ (yellow) take place.

\item[FIG. 11.4] (a) Projections (red circles) of the spiraling DNLs on the $2$D projected BZ (blue hexagon), where the portions of DNLs inside (solid) and outside (dotted) are shown. To sum up, there are two inequivalent DNLs in the $3$D rhombohedral BZ. The arrows indicate the spiraling senses in the increase of $k_{z}$. (b) $3$D rhombohedral BZ (red), with the unfilled dots on the high-symmetry points, in company with the $2$D projected BZ.

\item[FIG. 11.5]   Magnified LL spectra of RG at $B_{0}=30$ T for (a) $|E(n,B_{0},k_{z})|\leq0.12$, (b) $0.195\leq|E(n,B_{0},k_{z})|\leq0.255$ and (c) $0.335\leq|E(n,B_{0},k_{z})|\leq0.365$ (in the unit of $t$), black: minimal model; red: Onsager quantization; blue: numerical result.

\item[FIG. 11.6]   LL wave functions for $k_{z}=\pi/6d$ in RG at $B_{0}=20$ T. (a) One of the two degenerate set of LSs, plotted for $n=0$ and unoccupied $n=1$, $2$ and $3$. (b) The other set plotted for $n=0$ and unoccupied $n=1$. All the ordinate tick marks are labeled at $0$ and $0.005$.

\item[FIG. 11.7] (a) Spin polarization around the DP on the surface of TRS protected TI, where electron spin is locked to its momentum. (b) Electron spin degeneracy around a DP in the bulk of graphene.

\item[FIG. 11.8] Schematics for interpreting TRS protected $3$D strong TI. (a) LL spectrum when the continuous CS holds. (b) Hall plateaus arising from (a). (c) LL spectrum when the continuous CS is broken due to magnetic doping. (d) Hall plateaus with magnetic doping.

\end{itemize}

\newpage
\begin{figure}[p]
\centering \includegraphics[scale=0.8]{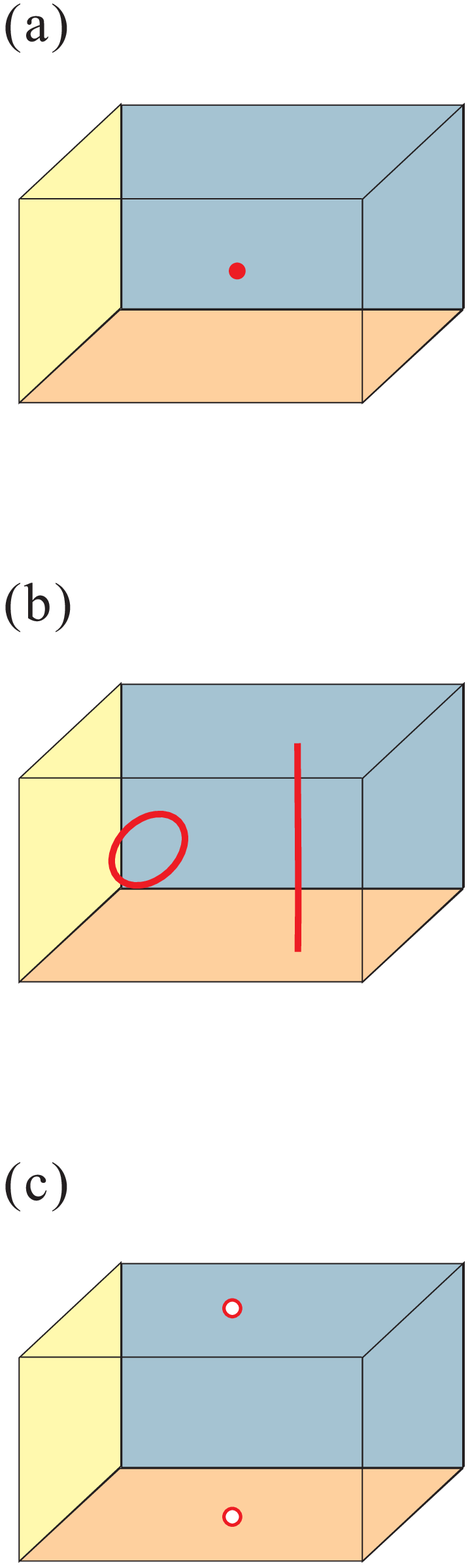}
\end{figure}

\begin{figure}[p]
\centering \includegraphics[scale=0.8]{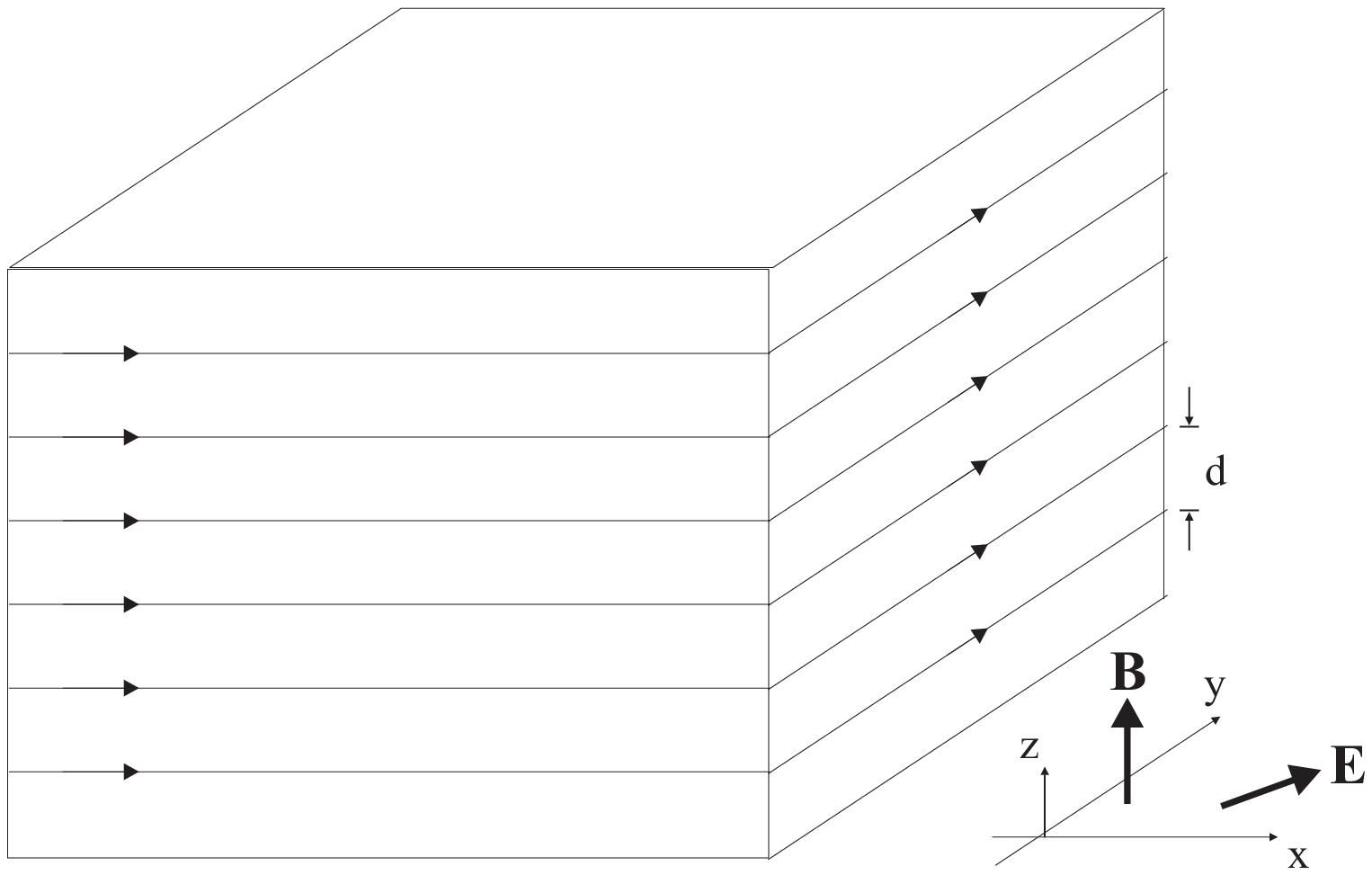}
\end{figure}

\begin{figure}[p]
\centering \includegraphics[scale=0.8]{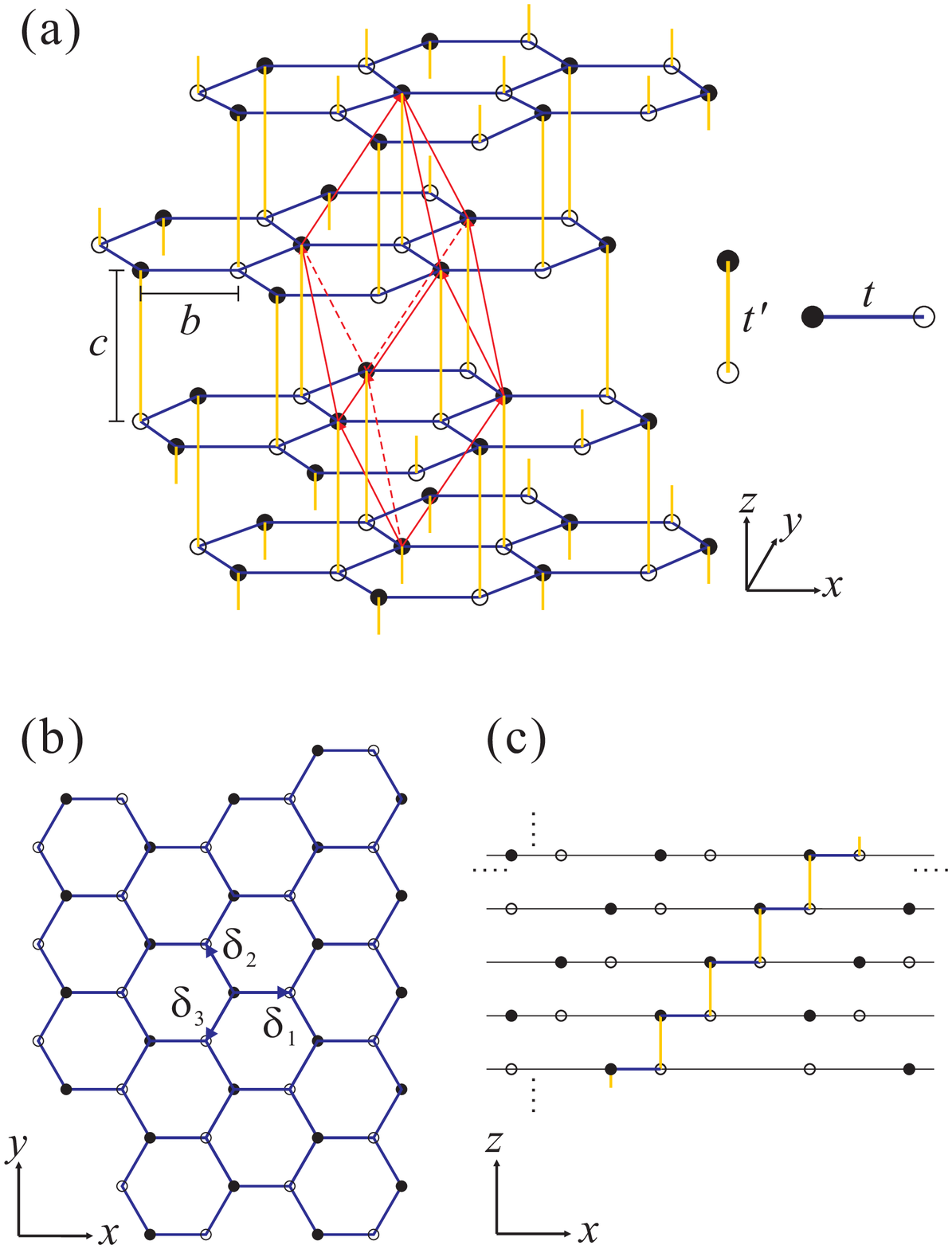}
\end{figure}

\begin{figure}[p]
\centering \includegraphics[scale=0.8]{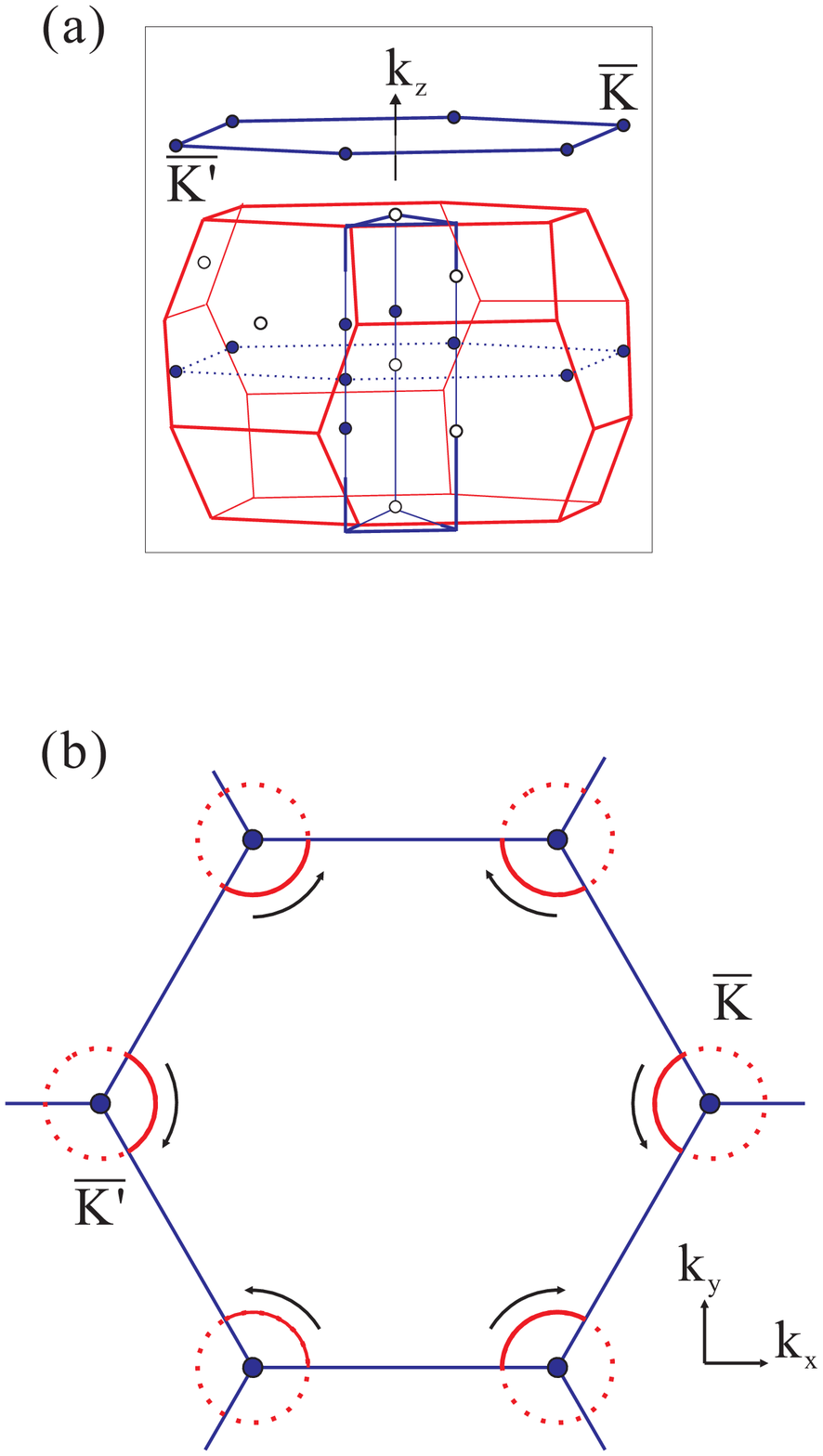}
\end{figure}

\begin{figure}[p]
\centering \includegraphics[scale=0.8]{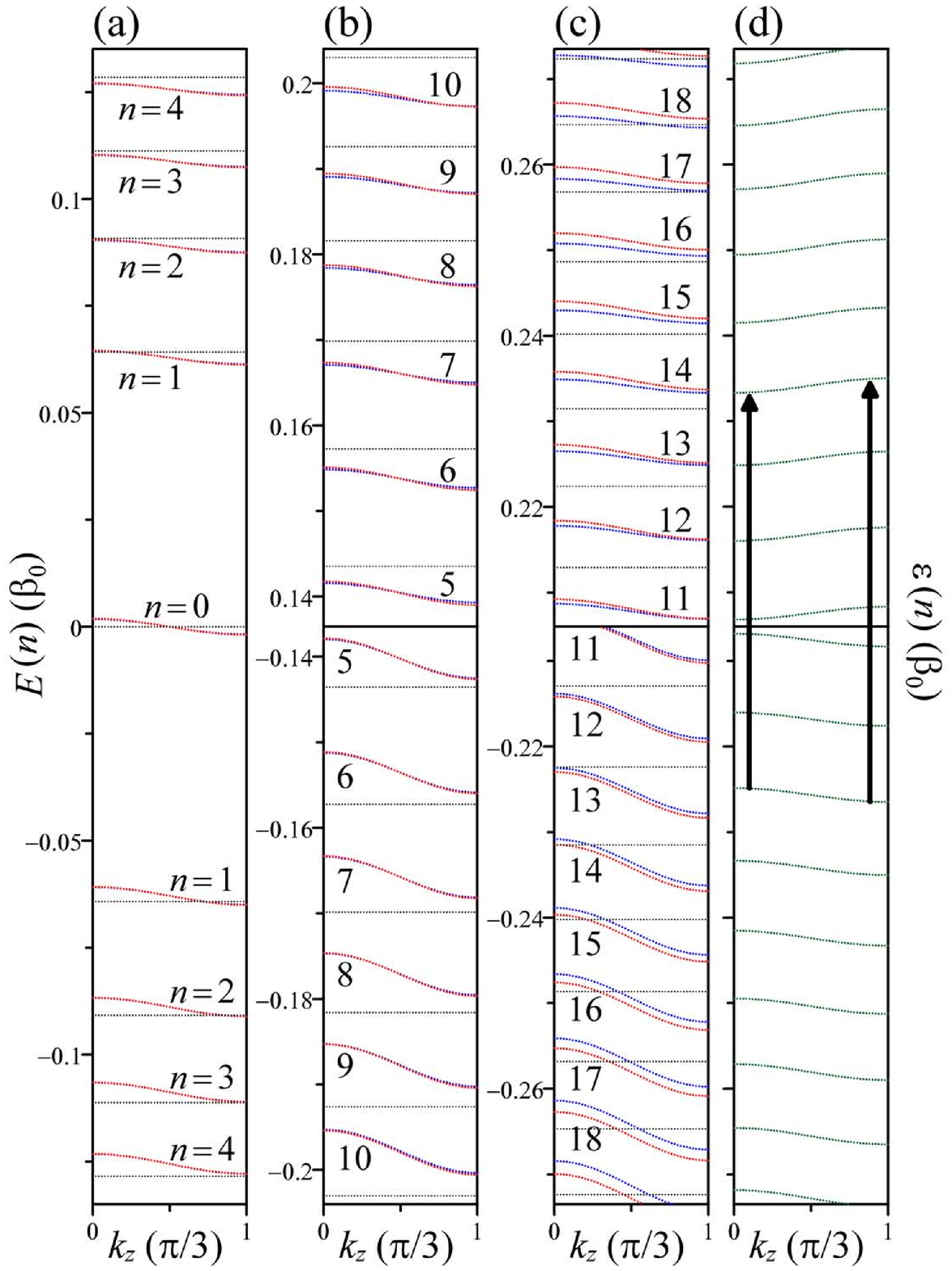}
\end{figure}

\begin{figure}[p]
\centering \includegraphics[scale=0.8]{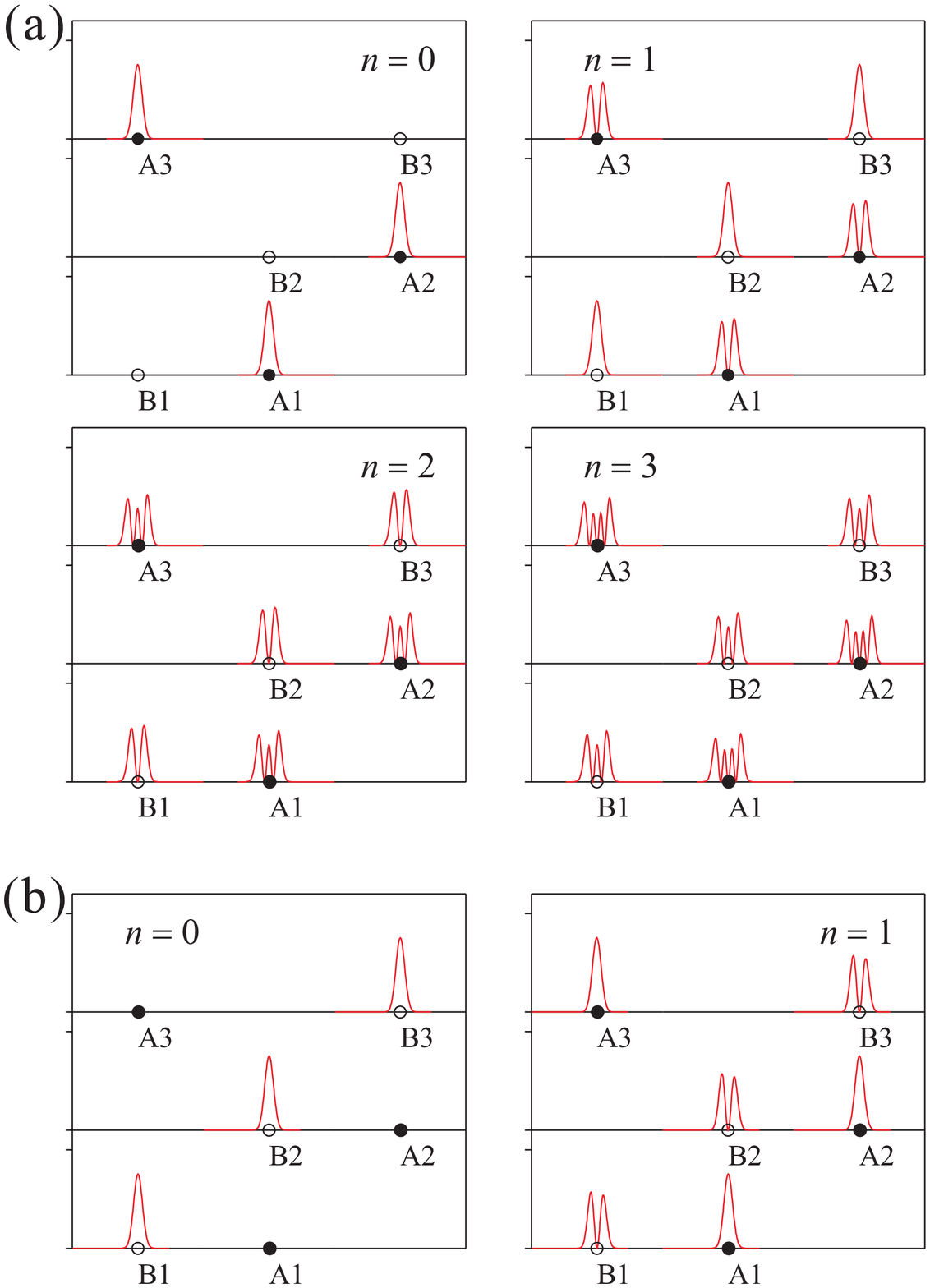}
\end{figure}

\begin{figure}[p]
\raggedleft \includegraphics[scale=0.8]{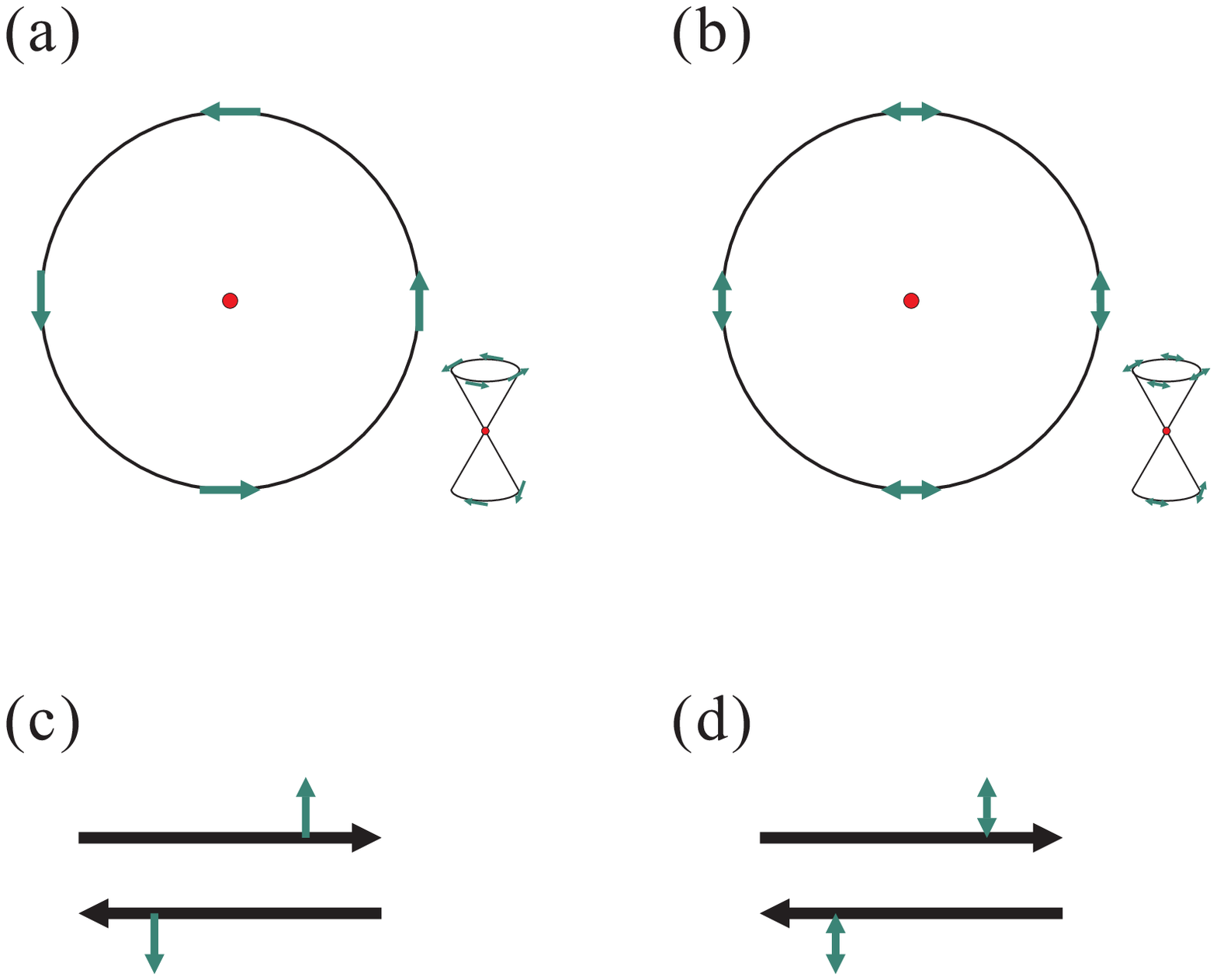}
\end{figure}

\begin{figure}[p]
\raggedleft \includegraphics[scale=0.8]{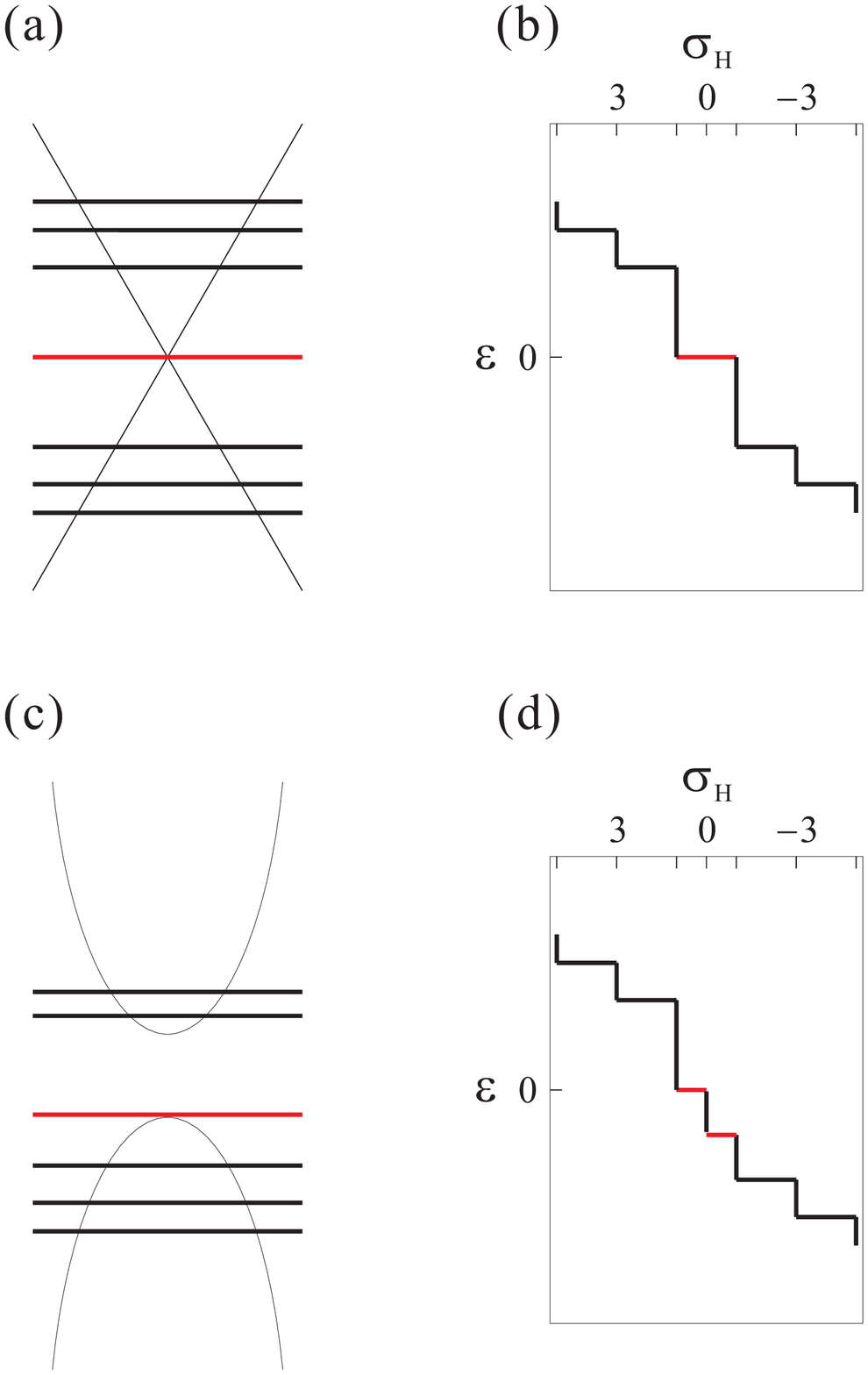}
\end{figure}

\end{document}